\documentclass[10pt]{article}
\usepackage{amsmath,amssymb,amsthm}
\usepackage{color}
\usepackage[nohead,margin=1.0in]{geometry}
\usepackage{graphicx,caption,subcaption}

\usepackage{amsfonts}

\usepackage{booktabs}

\usepackage{xcolor}

\newcommand{\N}{\mathbb{N}}
\newcommand{\R}{\mathbb{R}}

\newcommand{\myEqref}[1]{(\ref{#1})}

\newtheorem{remark}{Remark}[section]
\title{Mathematical models for magnetic particle imaging}
\author{Tobias Kluth\thanks{Center for Industrial Mathematics, University of Bremen,
Bibliothekstr. 5, 28357 Bremen, Germany (\texttt{tkluth@math.uni-bremen.de})}}
\begin{document}
\maketitle

\begin{abstract}
Magnetic particle imaging (MPI) is a relatively new imaging modality. 
The nonlinear magnetization behavior of nanoparticles in an applied magnetic field is employed to reconstruct an image of the concentration of nanoparticles.
Finding a sufficiently accurate model for the particle behavior is still an open problem. 
For this reason the reconstruction is still computed using a measured forward operator which is obtained in a time-consuming calibration process. 
The state of the art model used for the imaging methodology and first model-based reconstructions relies on strong model simplifications which turned out to cause too large modeling errors.
Neglecting particle-particle interactions, the forward operator can be expressed by a Fredholm integral operator of the first kind describing the inverse problem.
In this article we give an overview of relevant mathematical models which have not been investigated theoretically in the context of inverse problems yet. 
We consider deterministic models which are based on the physical behavior including relaxation mechanisms affecting the particle magnetization.
The behavior of the models is illustrated with numerical simulations for monodisperse as well as polydisperse tracer.
We further motivate linear and nonlinear problems beyond the solely concentration reconstruction related to applications. 
This model survey complements a recent topical review on MPI \cite{Knopp2017} and builds the basis for upcoming theoretical as well as empirical investigations.

\end{abstract}

\section{Introduction}

Magnetic particle imaging (MPI) is a relatively new imaging modality \cite{Gleich2005} which relies on the behavior of superparamagnetic iron oxide nanoparticles.  
The main goal is to reconstruct the spatially dependent concentration of the particles. 
Measurements are obtained from multiple receive coils where a potential is induced by the particle's nonlinear response to the applied dynamic magnetic field. 
These potential measurements are used for image reconstruction.
A high temporal resolution and a potentially high spatial resolution make MPI suitable for several in-vivo applications without the need for harmful radiation.  
Applications also benefit from the fast data acquisition of MPI.

The number of potential medical applications is still increasing. 
First proposed medical applications are vascular imaging and medical instrument tracking. 
The potential of imaging blood flow was shown first in {\it in vivo} experiments using a mouse \cite{weizenecker2009three}. 
The usability of a circulating tracer for long term monitoring has been investigated recently \cite{khandhar2017evaluation}.
Another medical application which benefits from the high temporal resolution is tracking medical instruments \cite{haegele2012magnetic}.
Recently it was shown that MPI is also suitable for tracking and guiding instruments for angioplasty \cite{Salamon:2016cz}.
Further promising applications of MPI can be found in cancer detection \cite{Yu2017} and cancer treatment by hyperthermia \cite{murase2015usefulness}.

The relationship between particle concentration and measured potential is modeled by a Fredholm integral equation of the first kind which is motivated by the suppression of particle interactions due to the nonmagnetic coating.
Determining the concentration is thus a linear inverse problem.
The integral kernel is defined by the properties of the receive coils, the applied magnetic fields and the dynamic behavior of the tracer materials.
MPI imaging methodologies are characterized by the applied magnetic fields which are generated by moving a field free point (FFP) \cite{Gleich2005} or a field free line (FFL) \cite{Weizenecker2008ffl} along a given trajectory rapidly.  
The most prominent FFP trajectories are Lissajous and Cartesian trajectories but also other trajectories were investigated in simulation studies \cite{Knopp2009trajectory}.
Cartesian sequences \cite{croft2012relaxation,Konkle2015a, Zheng2015} allow modeling the MPI signal by a spatial convolution.
In this case deconvolution methods are commonly used to obtain image reconstructions, also known as x-space reconstruction \cite{goodwill2010x, GoodwillConolly2011}.
In contrast, using a Lissajous trajectory, measurements can be obtained faster but the corresponding system matrix obeys a complex structure.
In the absence of suitable models, the system matrix is usually measured \cite{Knopp2009a, Gruettner_etal2013} for different scanner setups and tracer materials in a time-consuming measurement process where a ``delta'' probe is moved through the field of view. 
Besides the time-consuming measurement process the methodology suffers from high memory requirements particularly for three-dimensional imaging.  

The problem of modeling MPI, respectively finding the correct integral kernel, also known as system function, is an unsolved problem.
Existing model-based reconstructions incorporate particle behavior based on the theory of paramagnetism \cite{Knopp2009a,Knopp2010d,goodwill2010x,maerz2015modelbased,Kluth2017}.
Methods based on ideal magnetic fields \cite{rahmer2009signal,goodwill2010x} and on realistic magnetic fields \cite{Knopp2010d} are promising but they are not yet able to reach the quality of measured system functions. 
One possible reason may be a nonlinear dependence on the concentration which have been reported for large concentrations \cite{loewa2016}. 
But as the affected concentrations are larger than the concentrations commonly used in MPI experiments, these effects are not included in existing models used for imaging so far.
Another possible reason which arouses increasing interest in MPI is the particle relaxation which is likely to emerge for rapidly changing applied magnetic fields in MPI. 
Simplified models given by ordinary differential equations are combined with the superparamagnetic particle behavior \cite{croft2012relaxation} to deal with this issue.
The authors report a fitted relaxation time of $2.9 \ \mu$s for Resovist in an applied field of $22.9 \text{kHz}$ which causes blurred reconstructions when neglected in the reconstruction.
The dynamic behavior of the particles is affected by Brownian and N\'{e}el relaxation mechanisms \cite{rogge2013simulation,reeves2014approaches}.
For Resovist it was found that the behavior is mainly determined by N\'{e}el relaxation for immobilized particles while Brownian relaxation may influences the behavior in Resovist suspension depending on the frequency of the applied field \cite{ludwig2013characterization}.   
The still insufficient model in MPI, which neglects particle relaxation, motivates the increasing number of studies considering MPI excitation patterns in the context of Brownian and N\'{e}el relaxation \cite{weizenecker2012micro,Martens2013,Deissler2014,Shah2015,graeser2015trajectory}.
An experimental validation of the Brownian relaxation model using low frequencies in one-dimensional sinusodial excitation patterns \cite{Martens2013} emphasize its relevance. 
Relaxation in MPI applied magnetic fields has also been shown to be relevant in terms of a distinction of different kinds of tracers \cite{rahmer2015first}. 
Extending the particle models is of particular interest \cite{graeser2015trajectory} but have not been applied extensively to the imaging problem.
Due to the numerical challenges, the particle's physical dynamics have not been used for multidimensional image reconstructions so far. 
There only exist a limited number of works considering simplified models for particle relaxation while reconstructing the concentration \cite{croft2012relaxation} which can be extended easily to the multidimensional case.

Reconstructing the concentration in MPI is a linear ill-posed inverse problem \cite{Knopp_etal2008fc, maerz2015modelbased} which is solved commonly by applying Tikhonov regularization \cite{weizenecker2009three, Knopp2010e, rahmer2012analysis, Lampe_Fast_2012}.
Due to the complexity, the use of iterative solvers is advantageous.
In the literature the problem is solved preferably by using the algebraic reconstruction technique  \cite{weizenecker2009three, Knopp2010e,  Kaul2015, Knopp_Online_2016} combined with a nonnegativity constraint \cite{weizenecker2009three}.
Tikhonov regularization was applied to both kinds of system functions, data-based as well as model-based. 
Regularization techniques like fused lasso regularization or other gradient-based methods were recently applied to the data-based MPI problem \cite{storath2016edge, Konkle2015a}.
Directional total variation to incorporate a priori information from another modality (e.g. MRI) was also applied to the MPI problem in a first simulation study \cite{Bathke2017}.
An investigation of the current state of the art model with respect to operator uncertainty was done by using the total-least-squares approach which was combined with a standard Tikhonov penalty term as well as a sparsity promoting penalty term \cite{Kluth2017}.
Analogous to previous results in other applications \cite{gehre2012sparsity,gehre2014sparse} introducing sparsity improves the image reconstructions.
But a rigorous mathematical consideration of the parameter identification problems \cite{jin2012sparsity} in MPI is still missing.   
A first step in the direction of a theoretical understanding of the problem can be found in \cite{maerz2015modelbased} where the authors consider a similar problem by excluding the temporal nature of the MPI problem.

However, the use of mathematical models for imaging in MPI is very limited but it is highly desirable from various point of views. 
The potential medical applications benefit from the high temporal resolution of MPI measurements which requires computationally efficient reconstruction methods.
Particularly for three-dimensional imaging the development of fast model adapted algorithms is advantageous.   
The problem of time- and memory requirements in the system calibration may be solved in a more efficient way by taking a sufficient model into account.
Model-based reconstruction in MPI is still an open challenge which requires to identify a suitable mathematical model.
Furthermore, the problem of MPI for different applied magnetic field patterns have not been analyzed analytically in the context of inverse problems so far.
In this article we give an overview of different physical models suitable and proposed for MPI to build the basis for further investigations.
The overview is complemented by numerical illustrations of the model behavior. 
In Section \ref{sec:models} we formulate different models in a comprehensive mathematical way for monodisperse as well as polydisperse tracers.
Here we take into account the particle behavior in equilibrium, Brownian rotation, and N\'{e}el relaxation.
Additional information about the models and their relationship is provided in Appendices A, B, and C. 
To illustrate the behavior of the models, numerical simulations of the model behavior are shown in Section \ref{sec:numerics}.
The article concludes with a discussion in Section \ref{sec:discussion}.

\section{Preliminaries on magnetic particle imaging}
\label{sec:models}
This sections aims at technical definitions required for the formulation of MPI models. 
For a more detailed introduction to the principles of MPI the interested reader is referred to \cite{Knopp2012r}.
A recent review of MPI tracer development can be found in \cite{bauer2015magnetic} and for a detailed chronological overview on the technical developments in MPI during the last decade we refer to \cite{Knopp2017}.

\subsection{Definitions and notations}
The inherent nature of the problem is three-dimensional which is the reason why the relevant vector valued functions remain three dimensional even if the domain of the spatial variable is a subset of a $d$-dimensional affine subspace $E_d\subset \R^3$. 
Let $\Omega \subset E_d$, $d=1,2,3$, be a bounded domain in $E_d$ and let $S=\{ x\in \R^3 | \|x\|=1 \}$ be the unit sphere. 
Furthermore, let  $T>0$ denote the maximal data acquisition time and $I=[0,T]$ the time interval during which the measurement process takes place.
The temporal partial derivative of a function $g: I \rightarrow \R^k$, $k\in \N$, is denoted by $\dot{g}$, i.e., $\dot{g}=\frac{\partial}{\partial t} g$.

\subsection{Preliminaries}
In the following we define the problem and subsequently describe how the lower dimensional case is constructed.
The signal $v_k : I \rightarrow \R$, $k=1,\hdots,L$, which is obtained from the $L\in \N$ receive coils, is given by 
\begin{equation}
\label{eq:complete-problem}
 v_k (t) =- \int_I \int_\Omega c(x) \tilde{a}_k(t-\tau) s_k(x,t) \ dx \ d \tau -  \underset{=v_{\mathrm{E},k}(t)}{\underbrace{\int_I \int_{\R^3} \tilde{a}_k(t-\tau) p_k(x)^T \dot{ B}_\mathrm{app}(x,t)  \ dx \ d \tau}}
\end{equation}
where $c: \Omega \rightarrow \R^+ \cup \{0\}$ is the concentration of the magnetic nanoparticles and $s_k:\Omega \times I \rightarrow \R$, $k=1,\hdots,L$, denote the system functions characterizing the behavior of the nanoparticles.
$a_k: I \rightarrow \R$ , $k=1,\hdots,L$, are the periodic kernel functions of the analog filter in the signal acquisition chain and $\tilde{a}_k$ denotes its periodic continuation. 
$p_k: \R^3 \rightarrow \R^3$, $k=1,\hdots, L$, denotes the vector field which characterizes the sensitivity profile of the receive coils. 
In the remainder of the article it is assumed that the magnetic flux density $B_\mathrm{app}: \R^3 \times I \rightarrow \R^3$ of the applied magnetic field and the kernel functions $a_k$, $k=1,\dots,L$, are chosen in a way such that $v_{\mathrm{E},k}=0$ holds for all excitation signals $v_{\mathrm{E},k}: I \rightarrow \R$, $k=1,\hdots,L$, as defined in \myEqref{eq:complete-problem}.  
\begin{remark}
 Common choices for the analog filters $a_k$, $k=1,\hdots,L$, are band stop filters adapted to the frequencies of sinusodial excitations used in the subsequently described drive field.   
 The assumption regarding the excitation signals $v_{\mathrm{E},k}$, $k=1,\hdots,L$, is commonly made when the structure of the system functions is studied but it is not fulfilled in MPI applications in general \cite{Them2016}. 
 Efforts are made to remove the excitation signal or reduce its influence on the concentration reconstruction \cite{Knopp2010e,Weber2015z,Them2016,Kluth2017}. 
\end{remark}

The applied magnetic fields used in MPI can be characterized ideally by a spatially dependent magnetic field $g:\R^3 \rightarrow \R^3$ and a time-dependent homogeneous magnetic field $h:I \rightarrow \R^3$.
The applied magnetic field is then given by their superposition, i.e., $B_\mathrm{app}(x,t) = g(x) + h(t)$.
The ideal case relies on the assumption that the sensitivity profile of the field generating coils is homogeneous, respectively linear, in the field of view. 
The field $g$, also known as {\it selection field}, guarantees that a field-free-region is generated. 
Ideally, $g$ is assumed to be linear such that it can be represented by its transformation matrix $G\in \R^{3\times 3}$. 
Here two cases for the methodology are distinguished, namely whether a FFP is generated ($\mathrm{rank}( G) =3$) or a FFL is used ($\mathrm{rank}( G) =2$). 
The field $h$, also known as {\it drive field}, then moves the field-free-region along a certain trajectory.

\begin{remark}
 In the literature it was also proposed for the FFL approach to rotate the selection field over time such that the FFL is rotated \cite{Weizenecker2008ffl,Knopp2011FFLfourierslice}.
 In \cite{Knopp2011FFLfourierslice} the authors also show a relation between the Radon transform and the FFL approach combined with the particle model based on the Langevin function (subsequently termed {\it equilibrium model}).
 The selection field is then given by $g: \R^3 \times I \rightarrow \R^3$ with $g(x,t)=P(t)^TGP(t)x$ where $P:I\rightarrow \R^{3\times 3}$ is a rotation matrix for all $t\in I$.
\end{remark}

\begin{remark}
 In common MPI applications the drive field is defined by weighted sine or cosine functions with different frequencies in each component of $h$. 
 But also other kinds of excitation signals were investigated in simulations \cite{Knopp2009trajectory}.
\end{remark}

We now can formulate the most general version of the MPI problem.
It is then to find the concentration $c$ which fulfills 
\begin{equation}
\label{eq:general-problem}
\left\{
\begin{split}
 v_k(t)& = - \int_I \int_\Omega c(x) \tilde{a}_k(t-t') s_k(x,t') \ dx \ d t' \\
s_k & = \mu_0 p_k^T \frac{d}{dt} \bar{m} 
\end{split}
\right.
\end{equation}
for $k=1, \hdots, L$ and where $\bar{m}:\Omega \times I \rightarrow \R^3$ is the mean magnetic moment vector of the nanoparticles.
The 1st equation in \myEqref{eq:general-problem} models the analog filter process by a convolution with respect to $t'$. 
The spatial integration describes the induction of a potential in the receive coil and is obtained from Faraday's law of induction combined with the law of reciprocity to obtain the sensitivity profile of the coils \cite{Knopp2012r}.
The 2nd equation comprises the sensitivity of the receive coil and the particle behavior in the applied magnetic field.

The $d$-dimensional case for $d<3$ is constructed by assuming that the concentration is a $\delta$-distribution with respect to the orthogonal complement of the affine subspace $E_d$.
Then the $d$-dimensional problem is constructed by assuming $c(x)= \tilde{c}(x_1) \delta(x_2)$ where $x=x_1+x_2$ with $x_1\in E_d$, $x_2 \in E_d^\perp$, and $\tilde{c}: \Omega \subset E_d \rightarrow \R^+ \cup \{0\}$.
The parametrization of $\Omega\subset E_d$ then allows reformulating the spatial integral in \myEqref{eq:general-problem} as an integral over an integration domain $\Omega_d \subset \R^d$.
Given the affine linear parametrization $\Gamma : \Omega_d \rightarrow \Omega$ we can consider the problem with respect to $c_d: \Omega_d \rightarrow \R^+ \cup \{0 \}$, $c_d(x)=\tilde{c}(\Gamma(x))$.
All other spatially dependent functions are then treated analogously. 

\begin{remark}
The linear dependence on the concentration $c$ is based on the assumption that the particle-particle interactions can be neglected. 
There is increasing evidence that demagnetization effects relying on these interactions can significantly influence the particle signal \cite{loewa2016}. 
The nonlinear dependence is not considered in this article and remains to be explored in future work.
\end{remark}

\section{Models}
Now we are able to formulate various MPI models for different particle behavior and tracer compositions.
For the subsequent considerations we assume single-domain particles which is reasonable for sufficiently small diameters of the ferromagnetic core \cite{coffey2012thermal}.
In this case each particle has a uniform magnetization for any applied magnetic field. 
Here the magnetic moment is then considered for the whole single-domain particle instead of single atoms.
In general the single-domain particles are not isotropic due to the particle's shape, internal stress, and the internal structure \cite{coffey2012thermal}.
In the following it is assumed that the single-domain particles have an uniaxial anisotropy.

\subsection{Monodisperse tracer}

For monodisperse tracers it is assumed that the tracer material consists of one single kind of nanoparticles only, and all nanoparticles have the same characteristic behavior.
Two possible mechanism of relaxation are taken into account for MPI following \cite{rogge2013simulation,reeves2014approaches}. 
Based on the previous assumptions on the particles it is assumed that each particle is equipped with its magnetic moment vector which follows the external magnetic field. 
The particle can change the orientation of its magnetic moment vector by a rotation of the whole particle, known as Brownian rotation \cite{Coffey1992}, and by an internal rotation, known as N\'{e}el relaxation \cite{Neel1953,Brown1963}. 
First, we formulate the simplified model based on the Langevin function, termed {\it equilibrium model}.
Then the general MPI problem is specified with respect to the relaxation mechanisms resulting in MPI models which have not been studied with respect to the imaging problem yet.

\subsubsection{Equilibrium model} \ \\

\noindent One of the most extensively studied models in MPI is based on the Langevin function. 
It is the only model which has been studied with respect to the imaging problem so far. 
The model is motivated by the assumptions that the applied magnetic fields are static and the particles are in equilibrium. 
These assumptions also motivate the term {\it equilibrium model} which is used in the following. 
Using these assumptions, it is assumed that the mean magnetic moment vector of the nanoparticles immediately follows the magnetic field, i.e.,  
\begin{equation}
 \bar{m}(x,t) = \mathcal{L}_\beta(\| B_\mathrm{app}(x,t)\|) \frac{B_\mathrm{app}(x,t)}{\| B_\mathrm{app}(x,t)\|} 
\end{equation}
where $\mathcal{L}_\beta: \R \rightarrow \R$ is given in terms of the Langevin function by
\begin{equation}
\label{eq:Langevin}
 \mathcal{L}_\beta(z) =  m_0 \left( \coth( \beta z) - \frac{1}{ \beta z} \right)
\end{equation}
for $m_0,\beta >0$.
The Langevin function determines the length of the mean magnetic moment vector in equilibrium and can be derived from the Brownian rotation model stated below by assuming a static magnetic field.
A more detailed description of this relationship can be found in \ref{app:FP-Equilibrium}.  
The final problem based on the Langevin function then is to obtain the concentration $c$ from the following system of equations: 
\begin{equation}
\label{eq:problem-mono-equilibrium}
\left\{
\begin{aligned}
 &v_k(t) = - \int_I \int_\Omega c(x) \tilde{a}_k(t-t') s_k(x,t') \ dx \ d t' & &  \\
& s_k  = \mu_0 p_k^T \left( \left[ \left( \frac{\mathcal{L}_\beta'(\| B_\mathrm{app} \|)}{\| B_\mathrm{app} \|^2} - \frac{\mathcal{L}_\beta(\| B_\mathrm{app} \|)}{\| B_\mathrm{app} \|^3} \right) B_\mathrm{app} B_\mathrm{app}^T + \frac{\mathcal{L}_\beta(\| B_\mathrm{app} \|)}{\| B_\mathrm{app} \|} I_3 \right] \dot{B}_\mathrm{app} \right) & &  
\end{aligned}
\right.
\end{equation}
for $k=1,\hdots,L$ and the identity matrix $I_3\in\R^{3\times 3}$.

\begin{remark}
\label{rem:param-equilibrium}
 $m_0$ and $\beta$ are determined by the saturation magnetization $M_{\mathrm{C}}$ of the core material, the volume of the particle's core $V_{\mathrm{C}}$, the temperature $T_\mathrm{B}$, and the Boltzmann constant $\kappa_\mathrm{B}$, i.e., $m_0=M_{\mathrm{C}}V_{\mathrm{C}}$ and $\beta=m_0/(\kappa_\mathrm{B} T_\mathrm{B})$. 
Assuming spherical particles, the influence of the particle diameter $D$ is given by $V_{\mathrm{C}}=1/6 \pi D^3$.
Note that $\beta$ also depends on $D$ as it depends on $m_0$.
For example, particles consisting of magnetite with a typical diameter of $30 \text{ nm}$ $(20 \text{ nm})$ at room temperature $293 \text{ K}$ are characterized by $\beta \approx 2.1/\mu_0 \times 10^{-3}$ $(0.6/\mu_0 \times 10^{-3})$.
\end{remark}

\begin{remark}
Under certain assumptions on $B_\mathrm{app}$, this problem can be formulated in terms of a spatial convolution evaluated along the trajectory of the field free point with a temporally changing convolution kernel \cite{maerz2015modelbased,Kluth2017}. 
A first theoretical investigation related to this model can be found in \cite{maerz2015modelbased}.
\end{remark}

\subsubsection{Brownian rotation}\ \\

\noindent Here we assume that each particle has a magnetic moment vector which changes its direction due to the rotation of the whole particle \cite{Coffey1992,reeves2014approaches}.
Including the Brownian rotation and thermal noise results in a Langevin equation from which a deterministic system of differential equations is derived to model the mean magnetic moment vector.
The dynamics of a particle's magnetic moment vector $\tilde{m}:\Omega \times I \rightarrow m_0S$ with $\| \tilde{m}\| =m_0$ is given by
\begin{equation}
 \frac{\partial}{\partial t} \tilde{m} =  \frac{\nu}{m_0} \left( \tilde{m} \times B_\mathrm{app} + \tilde{D} \Gamma \right) \times \tilde{m}
\end{equation}
with physical parameters $\nu,\tilde{D}>0$ and where thermal noise is taken into account by the white noise component $\Gamma$ with $\langle \Gamma_i(t)\rangle =0$, $\langle \Gamma_i(t_1) \Gamma_j(t_2) \rangle = \delta_{ij} \delta(t_1-t_2)$, for all $t,t_1,t_2>0$ and $i,j=1,2,3$.
$\langle \cdot \rangle$ denotes the expectation value of a random variable. 
In case of zero noise the magnetic moment vector is directly damped into the direction of the magnetic field. 
The Fokker-Planck equation for the probability density function $f: S \times \Omega \times I \rightarrow \R^+\cup \{0\}$ is used to derive the mean magnetic moment vector.
The problem then becomes
\begin{equation}
\label{eq:problem-mono-brown}
\left\{
\begin{aligned}
& v_k(t) = - \int_I \int_\Omega c(x) \tilde{a}_k(t-t') s_k(x,t') \ dx \ d t' & & k=1,\hdots,L \\
&s_k  = \mu_0 p_k^T \frac{\partial \bar{m} }{\partial t}  & & k=1,\hdots,L\\
& \bar{m}(x,t)= m_0 \int_S m f(m,x,t) \ dm  & & \text{in } \Omega \times I \\
&\frac{\partial f}{\partial t} = - \text{div}_m \left( \nu \left( B_\mathrm{app}-m (m \cdot B_\mathrm{app})\right) f - \frac{1 }{2\tau}   \nabla_m f \right) & & \text{in } S\times \Omega \times I \\
&\int_S f(m,x,t) \ dm =1  & & \text{in } \Omega \times I \\
&f(\cdot,x,0)=f_0  & &  \text{in }\Omega 
\end{aligned}
 \right.
\end{equation}
for $m_0,\nu,\tau >0$ and where $f_0: S \rightarrow \R^+ \cup \{0\}$ with $\int_S f_0 \ dm=1$ is the initial distribution function.  
The 1st and 2nd equations of \myEqref{eq:problem-mono-brown} describe the general problem as stated in \myEqref{eq:general-problem}.
The 3rd equation defines the mean of the magnetic moment vector in terms of the probability density function $f$.
The 4th equation is the differential equation for the probability density function which takes into account the Brownian rotation. 
The derivation of the Fokker-Planck equation starting from the Langevin equation can be found in \ref{app:FP-Brown}.  
The 5th equation guarantees that $f$ is a probability density function and the sixth equation is the initial condition. 

\begin{remark} 
\label{rem:param-brown}
  $m_0$, $\nu$, and $\tau$ are determined by the saturation magnetization $M_{\mathrm{C}}$ of the core material, the volume of the core $V_{\mathrm{C}}$, the hydrodynamic volume $V_\mathrm{H}$ of the particles, the dynamic viscosity $\eta$, the temperature $T_\mathrm{B}$, and the Boltzmann constant $\kappa_\mathrm{B}$, i.e., $m_0=M_{\mathrm{C}}V_{\mathrm{C}}$, $\nu=\frac{m_0}{6V_\mathrm{H} \eta}$, and $\tau=3 V_\mathrm{H} \eta / (\kappa_\mathrm{B} T_\mathrm{B})$. 
  $\tau$ is known as the relaxation time.
Assuming spherical particles, the particle diameter $D$ influences $V_\mathrm{C}$ as well as $V_\mathrm{H}$.
Thus $m_0$, $\nu$, and $\tau$ depend on the particle diameter.
An overview of common particle parameters which are obtained from \cite{Deissler2014} can be found in Table \ref{tab:sim-parameters}.  
\end{remark}

\begin{remark}
Possible further applications in MPI are motivated by multi-color MPI \cite{rahmer2015first} which also distinguishes different kinds of tracer by their relaxation behavior \cite{utkur2017relaxation}.
In this case the characteristic behavior encoded in multiple system matrices allow the distinction.
In the Brownian rotation model it corresponds to the simultaneous reconstruction of $c$ and a spatially dependent viscosity $\eta$.
As a result a nonlinear inverse problem must be solved.
 \end{remark}

\subsubsection{N\'{e}el Relaxation}\ \\

\noindent Assume that the particles are spatially blocked such that Brownian rotation is suppressed. 
Each particle has a magnetic moment vector which changes its direction due to the change of internal electric states \cite{Neel1953,Brown1963,reeves2014approaches}. 
Including the applied magnetic field, the particle's anisotropy, and thermal noise results in a Langevin equation based on the Landau-Lifshitz-Gilbert equation for the particle's magnetic moment vector $\tilde{m}:\Omega \times I \rightarrow m_0S$ with $\|\tilde{m}\|=m_0$ given by
\begin{equation}
 \frac{\partial}{\partial t} \tilde{m} = \tilde{\gamma}  \left( (B_\mathrm{eff} + \tilde{D} \Gamma ) \times \tilde{m} + \frac{\alpha}{m_0} (\tilde{m} \times (B_\mathrm{eff} + \tilde{D} \Gamma) ) \times \tilde{m}\right)
\end{equation}
with physical parameters $\tilde{\gamma},\alpha, \tilde{D} >0$ and where $\Gamma$ is a white noise component with $\langle \Gamma_i(t)\rangle =0$, $\langle \Gamma_i(t_1) \Gamma_j(t_2) \rangle = \delta_{ij} \delta(t_1-t_2)$, for all $t,t_1,t_2>0$ and $i,j=1,2,3$.
In contrast to Brownian rotation the magnetic moment vector moves on a precessional trajectory while it is damped in direction of the magnetic field.  
From this equation a deterministic system of differential equations is derived to model the mean magnetic moment vector $\bar{m}$.
In contrast to the previous models, an effective magnetic field $B_\mathrm{eff}:S \times \Omega \times I \rightarrow \R^3$ is considered which consists not only of the applied magnetic field.
Here the uniaxial anisotropy of the particles is modeled by a field $B_\mathrm{anis}:S \rightarrow \R^3$ with $B_\mathrm{anis}(m)=2 \frac{K_\mathrm{anis} V_\mathrm{C}}{m_0} (m \cdot n) n $ for a given anisotropy direction $n\in S$ and anisotropy constant $K_\mathrm{anis}\in \R$.
This field is obtained from the Stoner-Wohlfarth model for uniaxial anisotropic particles \cite{stoner1991mechanism,Tannous2008}.
Here $K_\mathrm{anis}>0$ corresponds to the desired case of particles having an easy axis while $K_\mathrm{anis}<0$ describes an easy plane.   
The effective magnetic field is then given by the superposition of applied and anisotropy field, i.e., $B_\mathrm{eff}= B_\mathrm{app} + B_\mathrm{anis}$.
The Fokker-Planck equation for the probability density function $f: S \times \Omega \times I \rightarrow \R^+\cup \{0\}$ is used to derive the mean magnetic moment vector.
The problem then becomes
\begin{equation}
\label{eq:problem-mono-neel}
\left\{
\begin{aligned}
& v_k(t) = - \int_I \int_\Omega c(x) \tilde{a}_k(t-t') s_k(x,t') \ dx \ d t' & & k=1,\hdots,L \\
&s_k  = \mu_0 p_k^T \frac{\partial \bar{m} }{\partial t}  & & k=1,\hdots,L\\
& \bar{m}(x,t)= m_0 \int_S m f(m,x,t) \ dm  & & \text{in } \Omega \times I \\
&\frac{\partial f}{\partial t} = - \mathrm{div}_m \left( \tilde{\gamma} \left( B_\mathrm{eff} \times m + \alpha ( B_\mathrm{eff} - (m\cdot B_\mathrm{eff}) m)\right) f - \frac{1}{2 \tau}  \nabla_m f \right) & & \text{in } S\times \Omega \times I \\
&\int_S f(m,x,t) \ dm =1  & & \text{in } \Omega \times I \\
&f(\cdot,x,0)=f_0  & &  \text{in }\Omega 
\end{aligned}
 \right.
\end{equation}
for $m_0,\tilde{\gamma},\alpha,\tau >0$ and where $f_0: S \rightarrow \R^+ \cup \{0\}$ with $\int_S f_0 \ dm=1$ is the initial distribution function.  
The 1st and 2nd equation of \myEqref{eq:problem-mono-neel} describe the general problem as stated in \myEqref{eq:general-problem}.
The 3rd equation defines the mean of the magnetic moment vector with respect to the probability density function $f$.
The 4th equation is the differential equation for the probability density function which takes into account the N\'{e}el relaxation. 
The effective magnetic field is $B_\mathrm{eff}= B_\mathrm{app} + 2 \frac{K_\mathrm{anis} V_\mathrm{C}}{m_0} (m \cdot n) n$, $K_\mathrm{anis},V_\mathrm{C}>0$ and $n\in S$.
Depending on the choice of $n\in S$, the anisotropy field may counteract the applied magnetic field.
The larger the applied magnetic field strength the weaker is the influence of the anisotropy field.
The derivation of the Fokker-Planck equation starting from the corresponding Langevin equation can be found in \ref{app:FP-Neel}.  
The 5th equation guarantees that $f$ is a probability density function and the sixth equation is the initial condition.  

\begin{remark} 
\label{rem:param-neel}
  $m_0$, $\tilde{\gamma}$, and $\tau$ are determined by the saturation magnetization $M_{\mathrm{C}}$ of the core material, the volume of the core $V_{\mathrm{C}}$, the gyromagnetic ratio $\gamma$, the damping parameter $\alpha$, the temperature $T_\mathrm{B}$, and the Boltzmann constant $\kappa_\mathrm{B}$, i.e., $m_0=M_{\mathrm{C}}V_{\mathrm{C}}$, $\tilde{\gamma}=\frac{\gamma}{1+\alpha^2}$, and relaxation time $\tau= \frac{m_0}{2 \alpha \tilde{\gamma} \kappa_\mathrm{B} T_\mathrm{B}}$. 
Assuming spherical particles, the particle diameter $D$ influences $V_\mathrm{C}$.
Thus $m_0$ and $\tau$ depend on the particle diameter.
Note that $B_\mathrm{anis}$ does not depend on the diameter as the volume of the core $V_\mathrm{C}$ cancels out.
An overview of common particle parameters which are obtained from \cite{Deissler2014} can be found in Table \ref{tab:sim-parameters}.

\end{remark}

\begin{remark}
The temperature influences the behavior of the nanoparticles. 
In the context of multi-color MPI it was also suggested to determine the temperature while reconstructing the concentration \cite{stehning2016simultaneous}.
The authors motivate the simultaneous reconstruction by real time monitoring in hyperthermia applications.
The problem of finding a spatially dependent temperature $T_\mathrm{B}$ has a nonlinear nature requiring a deeper analysis of the problem.
Simultaneous usage of different kinds of particles differing in their physical properties result in another possible nonlinear problem similar to the case of multi-color MPI with viscosity mapping. 
Structural differences might also be found in the relaxation times $\tau$ indicating different environmental structures, e.g., whether particles are blocked or non-blocked.
 \end{remark}

\subsection{Polydisperse tracer}

In MPI, polydisperse tracer has been proposed in context of the equilibrium model for reconstruction so far.
The investigations of tracer materials commonly used in MPI motivate the introduction of a distribution of the core diameter \cite{Knopp2009a} which is mainly motivated by related physical investigations \cite{Kiss1999,Eberbeck2011,Du2013,Khandhar2013}.
The tracer material is then modeled by a distribution of particles differing in their diameter $D>0$. 
Assuming that the particle's diameter distribution is given by a density function $\rho : \R^+ \rightarrow \R^+ \cup \{ 0 \}$ with $\| \rho \|_{L^1(\R^+)}=1$, the extended problems are obtained.
Below we state these models to highlight the parameter dependence on the particle's core diameter.
For the same reasons as for the monodisperse problems with Brownian rotation and N\'{e}el relaxation their polydisperse extensions have not been considered for imaging yet.
\begin{remark}
The distribution function of the particle diameters for polydisperse tracers can be approximated by a log-normal distribution \cite{Kiss1999,Eberbeck2011,Du2013,Khandhar2013}.  
For example, for Resovist a mean $\mu=\ln(13 \times 10^{-9})$ and a standard deviation of $\sigma=0.37$ of a log-normal distribution were reported for the diameter \cite{Eberbeck2011}, i.e.
$\rho(D)= \frac{1}{D\sigma \sqrt{2\pi}} e^{-(\ln (D)- \mu )^2 / (2 \sigma ) ^2 } $.
\end{remark}

The {\it equilibrium model} stated in \myEqref{eq:problem-mono-equilibrium} can be extended to polydisperse tracers by adapting the function defining the length of the mean magnetic moment vector in \myEqref{eq:Langevin}.
The extended problem is then given by
\begin{equation}
\left\{
\begin{aligned}
 &v_k(t) = - \int_I \int_\Omega c(x) \tilde{a}_k(t-t') s_k(x,t') \ dx \ d t' & &  \\
& s_k  = \mu_0 p_k^T \left( \left[ \left( \frac{\mathcal{L}_{\beta,\rho}'(\| B_\mathrm{app} \|)}{\| B_\mathrm{app} \|^2} - \frac{\mathcal{L}_{\beta,\rho}(\| B_\mathrm{app} \|)}{\| B_\mathrm{app} \|^3} \right) B_\mathrm{app} B_\mathrm{app}^T + \frac{\mathcal{L}_{\beta,\rho}(\| B_\mathrm{app} \|)}{\| B_\mathrm{app} \|} I_3 \right] \dot{B}_\mathrm{app} \right) & &  
\end{aligned}
\right.
\end{equation}
where $\mathcal{L}_{\beta,\rho}: \R \rightarrow \R$ is given in terms of the Langevin function by
\begin{equation}
 \mathcal{L}_{\beta,\rho}(z) = \int_{\R^+} \rho(D)  m_0(D) \left( \coth(\beta(D) z) - \frac{1}{ \beta(D) z} \right) \ dD
\end{equation}
for $m_0,\beta: \R^+  \rightarrow \R^+$ describing the influence of the particle diameter on the volume of the core, respectively the magnetic moment.
For further details on the physical paramters and their relationship to the particle diameter, we refer to Remark \ref{rem:param-equilibrium}.

The {\it Brownian rotation model} stated in \myEqref{eq:problem-mono-brown} can be extended for polydisperse tracers by extending the domain of the magnetic moment vector's probability density function $f: S\times \Omega \times I \times \R^+ \rightarrow \R^+ \cup \{ 0 \} $. 
The {\it Brownian rotation model} then becomes 
\begin{equation}
\label{eq:problem-poly-brown}
\left\{
\begin{aligned}
& v_k(t) = - \int_I \int_\Omega c(x) \tilde{a}_k(t-t') s_k(x,t') \ dx \ d t' & & k=1,\hdots,L \\
&s_k  = \mu_0 p_k^T \frac{\partial \bar{m} }{\partial t}  & & k=1,\hdots,L\\
& \bar{m}(x,t)= \int_{\R^+} \rho(D) m_0(D) \int_S m f(m,x,t,D) \ dm \ dD  & & \text{in } \Omega \times I \\
&\frac{\partial f}{\partial t} = - \text{div}_m \left( \nu(D) \left( B_\mathrm{app}-m (m \cdot B_\mathrm{app})\right) f - \frac{1 }{2\tau(D)}   \nabla_m f \right) & & \text{in } S\times \Omega \times I \times \R^+ \\
&\int_S f(m,x,t,D) \ dm =1  & & \text{in } \Omega \times I \times \R^+ \\
&f(\cdot,x,0,D)=f_0  & &  \text{in }\Omega \times \R^+ 
\end{aligned}
 \right.
\end{equation}
for $m_0,\nu,\tau: \R^+ \rightarrow \R^+$ and where $f_0: S \rightarrow \R^+ \cup \{0\}$ with $\int_S f_0 \ dm=1$. 
In contrast to the monodisperse model, the 3rd equation of \myEqref{eq:problem-poly-brown} comprises the mean of the mean magnetic moment vector and the mean over the particle diameter. 
The differential equation for the probability density function in the 4th, 5th, and 6th equation are extended to take the dependence on the particle diameter $D$ into account.
For further details on the physical parameters and their relation to the particle diameter we refer to Remark \ref{rem:param-brown}.

The {\it N\'{e}el relaxation model} is extended analogously by considering  $f: S\times \Omega \times I \times \R^+ \rightarrow \R^+ \cup \{ 0 \} $.
It is thus given by
\begin{equation}
\label{eq:problem-poly-neel}
\left\{
\begin{aligned}
& v_k(t) = - \int_I \int_\Omega c(x) \tilde{a}_k(t-t') s_k(x,t') \ dx \ d t' & & k=1,\hdots,L \\
&s_k  = \mu_0 p_k^T \frac{\partial \bar{m} }{\partial t}  & & k=1,\hdots,L\\
& \bar{m}(x,t)= \int_{\R^+} \rho(D) m_0(D) \int_S m f(m,x,t,D) \ dm \ dD & & \text{in } \Omega \times I \\
&\frac{\partial f}{\partial t} = & & \\
& - \mathrm{div}_m \left( \tilde{\gamma} \left( B_\mathrm{eff} \times m + \alpha ( B_\mathrm{eff} - (m\cdot B_\mathrm{eff}) m)\right) f - \frac{1}{2 \tau(D)}  \nabla_m f \right) & & \text{in } S\times \Omega \times I \times \R^+ \\
&\int_S f(m,x,t,D) \ dm =1  & & \text{in } \Omega \times I \times \R^+ \\
&f(\cdot,x,0,D)=f_0  & &  \text{in }\Omega \times \R^+ 
\end{aligned}
 \right.
\end{equation}
for $m_0,\tau: \R^+ \rightarrow \R^+$, $\tilde{\gamma},\alpha >0$ and where $f_0: S \rightarrow \R^+ \cup \{0\}$ with $\int_S f_0 \ dm=1$, cf. Remark \ref{rem:param-neel} for further details on the physical parameters and their dependence on the diameter $D$.  
The 3rd equation of \myEqref{eq:problem-poly-neel} comprises the mean of the mean magnetic moment vector and the mean over the particle diameter and the 4th, 5th, and 6th equation are extended to take the dependence on the particle diameter into account.

\begin{remark}
 The polydisperse models are parametric equations which are similar to the monodisperse models based on the stochastic differential equations.
\end{remark}

\section{Numerical examples}
\label{sec:numerics}
To illustrate the behavior of different models we computed the temporal derivative of the mean magnetic moment vector $\bar{m}$ in one special case where the probability density function is circular symmetric with respect to the $e_3$-axis.
On the $e_3$-axis the third component of the mean magnetic moment vector is nonzero only.
The numerical solution is based on the formulation of a system of ordinary differential equations obtained by approximating the probability density functions for Brownian and N\'{e}el relaxation by a finite number of Legendre polynomials.
Further details about the numerical solution which follows \cite{Deissler2014} can be found in \ref{app:FP-numerics1D}.
In the subsequent simulations the first $50$ Legendre polynomials were used.
The initial distribution $f_0$ is assumed to be uniform.

The simulation setup is as follows.
We assume an excitation in the direction of $e_3$ only and consider the case that $\Omega \subset E_d=\{ q e_3 | q \in \R\}$.
The drive field is given by $h(t)= A \cos(2 \pi f t) e_3$ where $A>0$ is the excitation amplitude and $f>0$ is the excitation frequency.
The selection field is assumed to be linear with a diagonal transformation matrix $G\in\R^{3\times 3}$. 

For the simulations we use physical parameters typical for MPI applications.
An overview of the parameters can be found in Table \ref{tab:sim-parameters}.
The remaining parameters are computed according to Remarks \ref{rem:param-equilibrium}, \ref{rem:param-brown}, and \ref{rem:param-neel}.

\begin{figure}
	 
 	\begin{subfigure}[t]{0.75\textwidth}
 		\includegraphics[width=\textwidth]{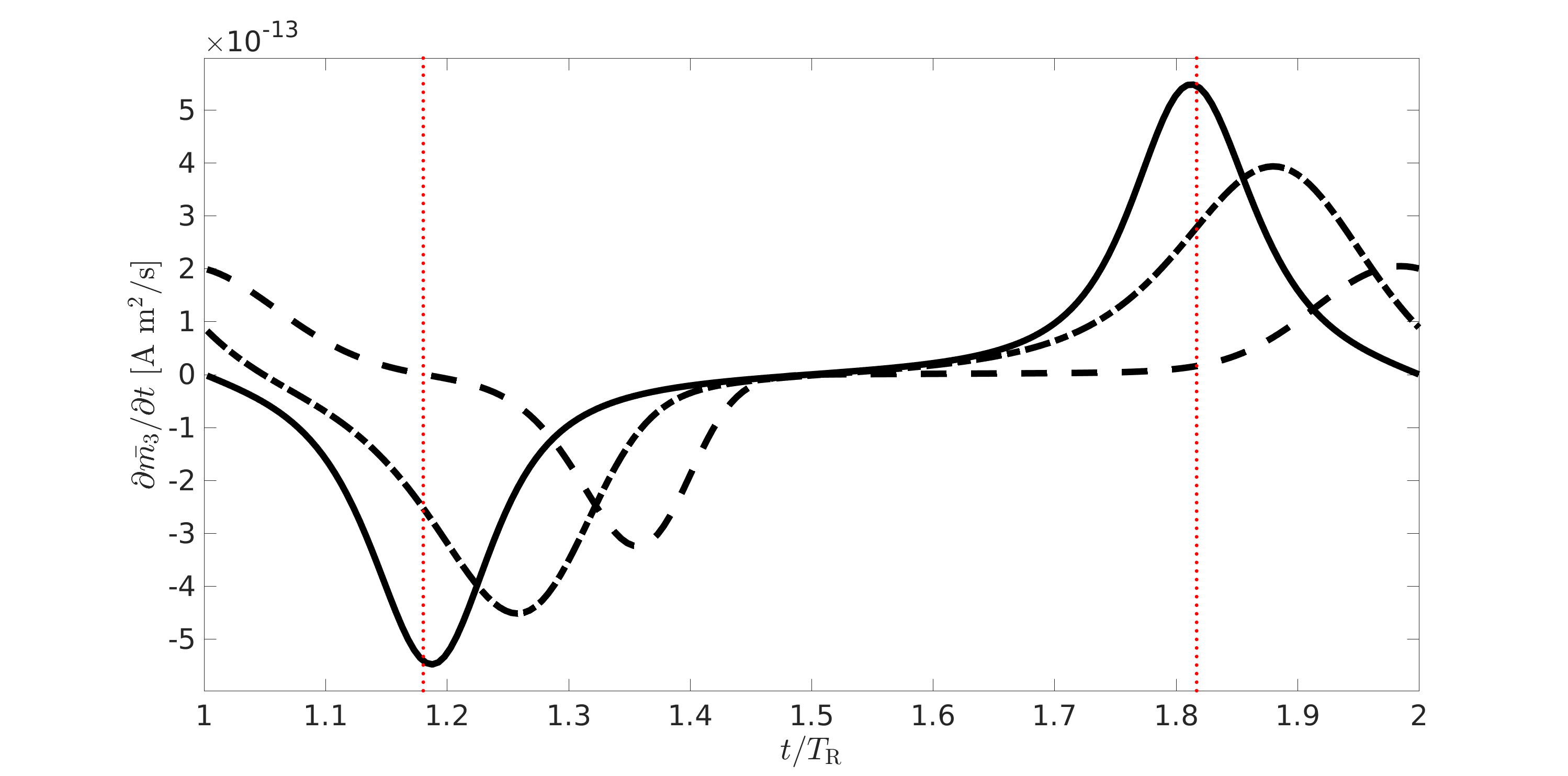}
		\caption{$z= -2.5 \text{ mm}$  }
		\label{subfig11}
	\end{subfigure}
	\begin{subfigure}[t]{0.75\textwidth}
 		\includegraphics[width=\textwidth]{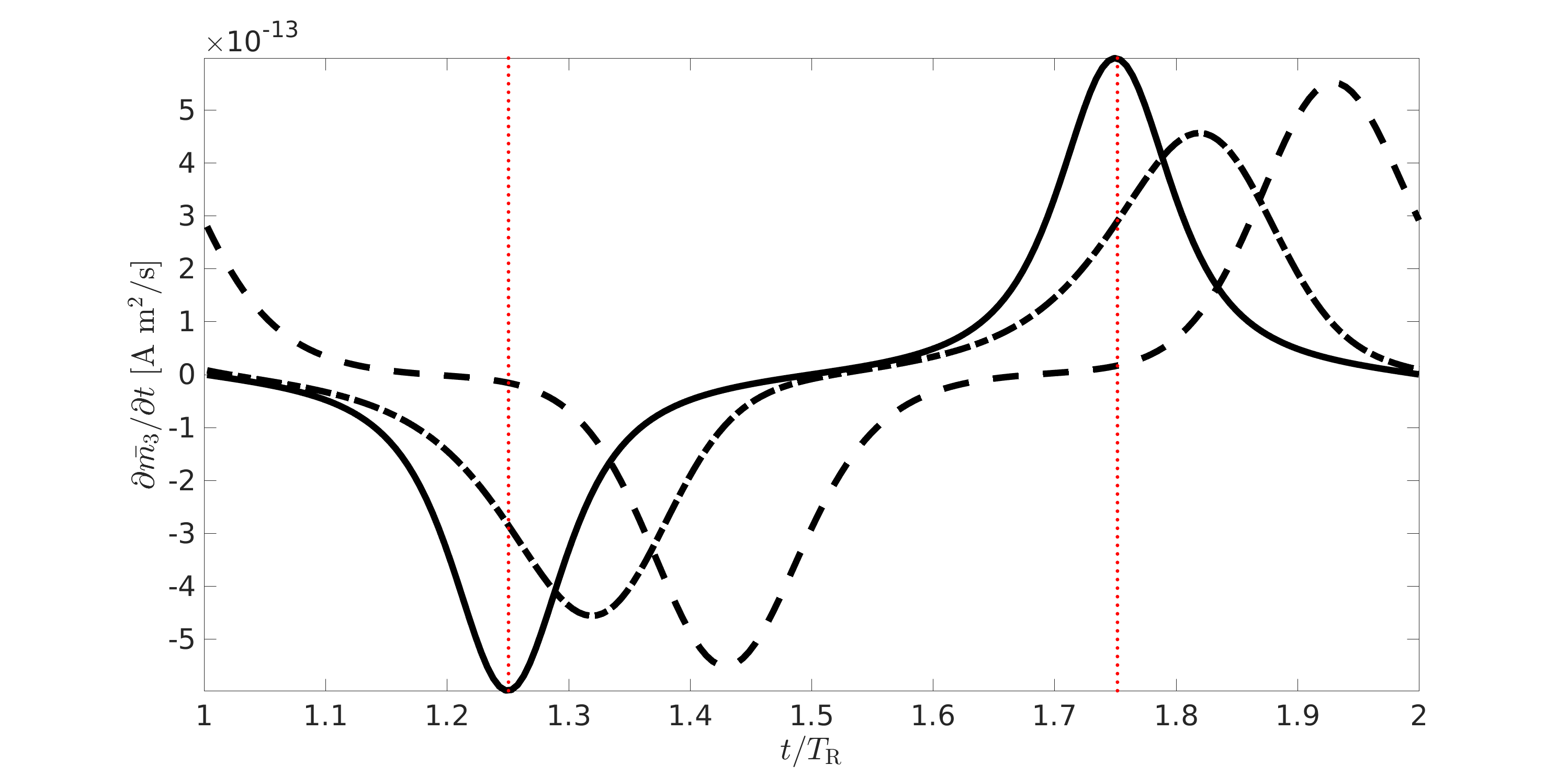}
		\caption{$z= 0 \text{ mm}$  }
		\label{subfig12}
	\end{subfigure}
	\begin{subfigure}[t]{0.75\textwidth}
 		\includegraphics[width=\textwidth]{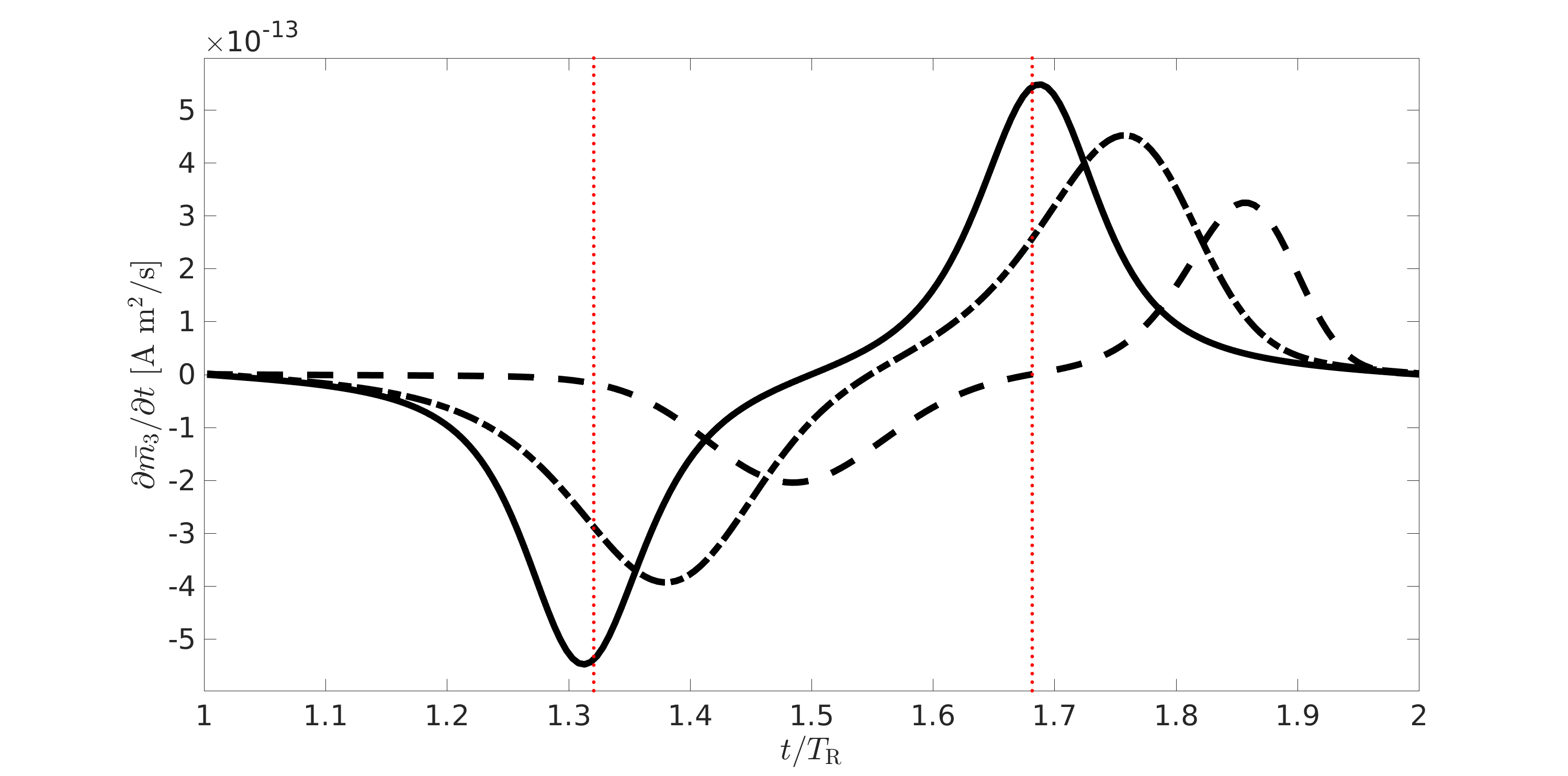}
		\caption{$z= 2.5 \text{ mm}$  }
		\label{subfig13}
	\end{subfigure}
		\begin{subfigure}[t]{0.15\textwidth}
 		\includegraphics[width=\textwidth]{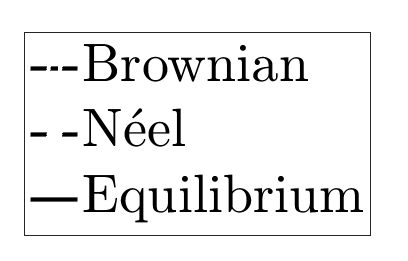}
		
	\end{subfigure}
	 
\caption{Simulated mean magnetic moment vector in $e_3$-direction considering the monodisperse models with parameters and particularly particle diameter specified in Table \ref{tab:sim-parameters}. The dotted vertical lines (red) highlight the point in time with zero applied magnetic field. }
\label{fig:model_simulations1D_mono}
\end{figure}

The simulations for the monodisperse models are illustrated in Figure \ref{fig:model_simulations1D_mono} for a diameter of $20 \text{ nm}$. 
The graphs of the equilibrium model are point symmetric to the point $(1.5,0)$ which is a direct result of the cosine excitation.
The extremal points in the equilibrium model are close to the field free point in time.
The Brownian rotation model shows damped and skewed behavior in time direction when it is compared to the equilibrium model.
This is due to the viscous rotation of the particles.
Neglecting the shift in time, the N\'{e}el relaxation shows a similar qualitative behavior in terms of temporal symmetry at the origin ($z=0\text{ mm}$) compared to the equilibrium model.
The shift in time may be related to the particle anisotropy with preferred $e_3$-direction. 
A certain strength of the magnetic field is required before the magnetization changes. 
More structural differences can be observed at the remaining $z$-positions where one extremal point is more damped than the other one.
The more it is damped the less is the applied magnetic field able to counteract the anisotropy in the preferred direction of anisotropy.

\begin{figure}
	 
 	\begin{subfigure}[t]{0.75\textwidth}
 		\includegraphics[width=\textwidth]{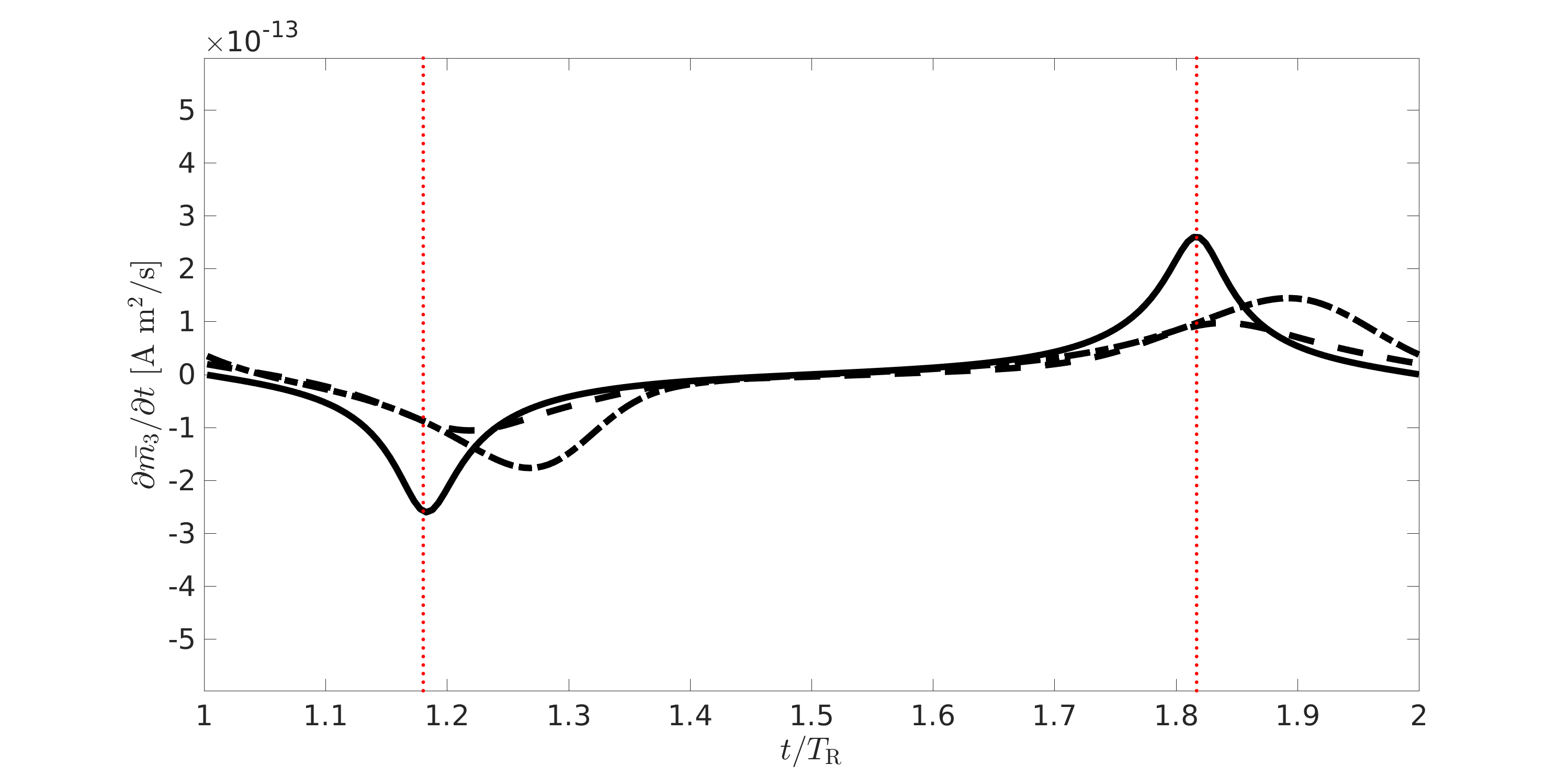}
		\caption{$z= -2.5 \text{ mm}$  }
		\label{subfig21}
	\end{subfigure}
 	\begin{subfigure}[t]{0.75\textwidth}
 		\includegraphics[width=\textwidth]{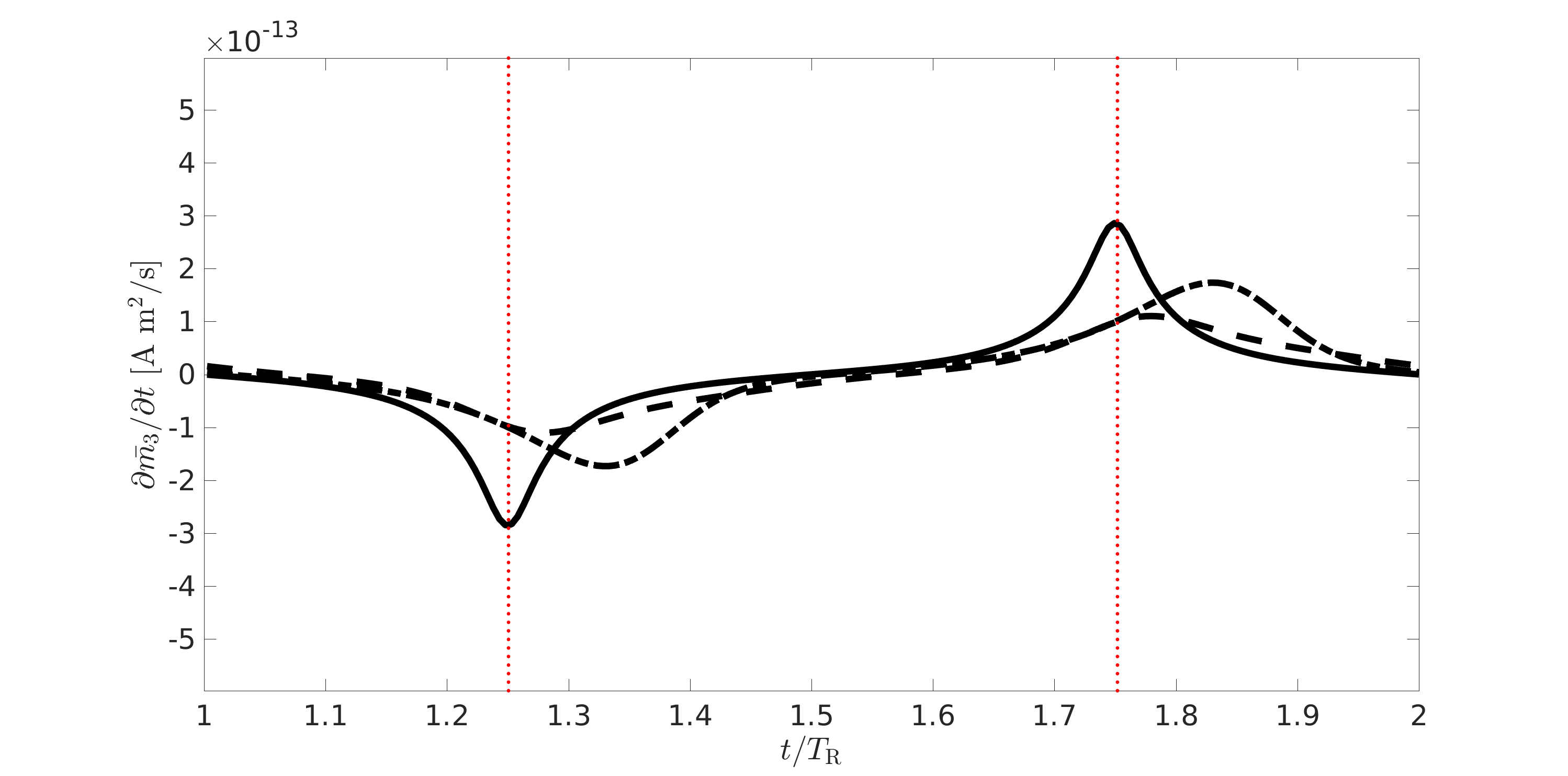}
		\caption{$z= 0 \text{ mm}$  }
		\label{subfig22}
	\end{subfigure}
 	\begin{subfigure}[t]{0.75\textwidth}
 		\includegraphics[width=\textwidth]{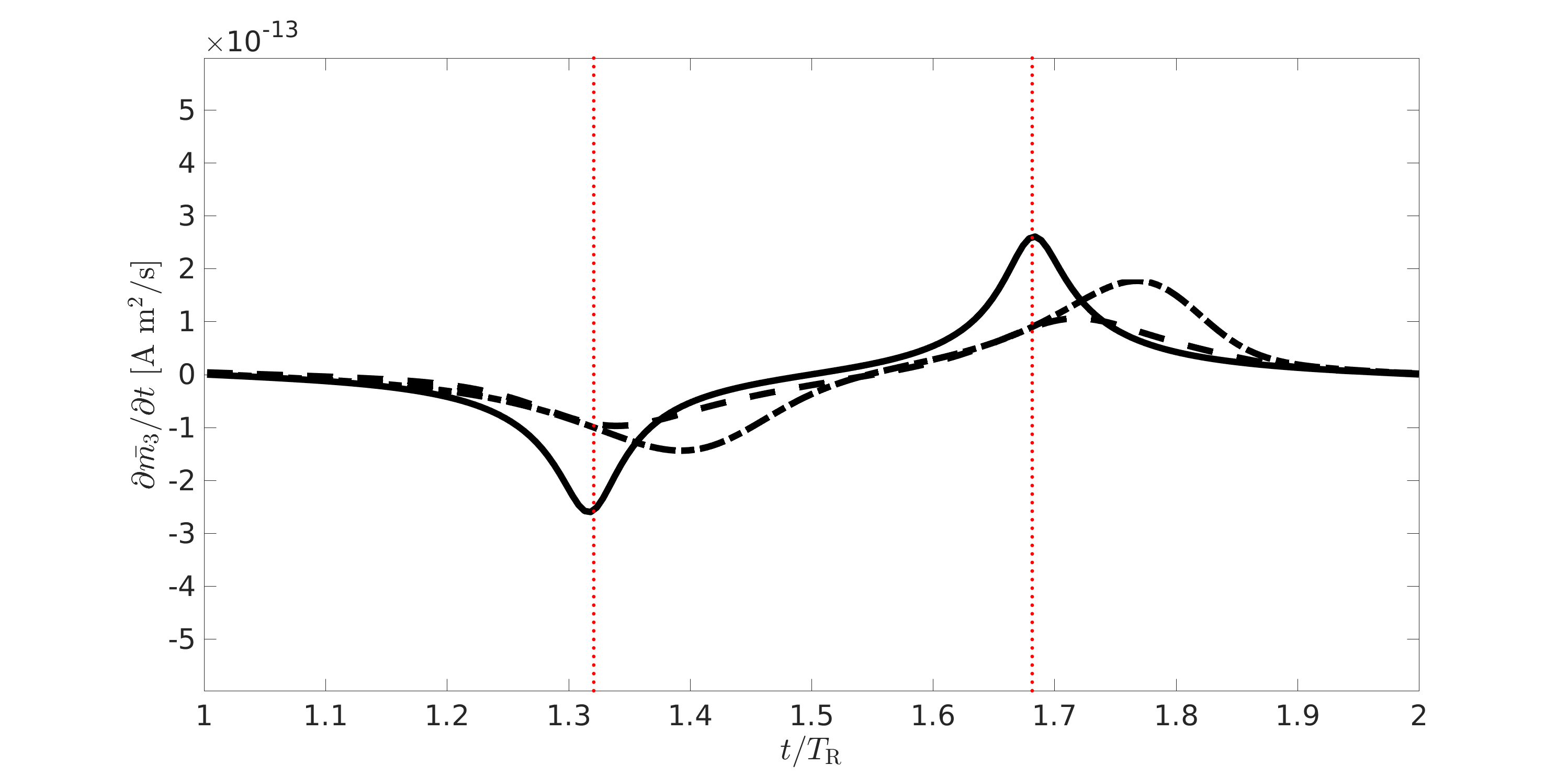}
		\caption{$z= 2.5 \text{ mm}$  }
		\label{subfig23}
	\end{subfigure}
			\begin{subfigure}[t]{0.15\textwidth}
 		\includegraphics[width=\textwidth]{legend_crop.png}
		
	\end{subfigure}
	
\caption{Simulated mean magnetic moment vector in $e_3$-direction considering the polydisperse models with parameters and particularly the lognormal particle diameter distribution specified in Table \ref{tab:sim-parameters}. The dotted vertical lines (red) highlight the point in time with zero applied magnetic field.}
\label{fig:model_simulations1D_poly}
\end{figure}

The simulations for the polydisperse models are illustrated in Figure \ref{fig:model_simulations1D_poly}.
The polydisperse equilibrium model is smaller in amplitude but the qualitative behavior remains similar to the monodisperse version with a diameter of $20$ nm. 
The change in amplitude is a direct result of the diameter distribution shifted towards particle diameters smaller than $20$ nm.
The relationship between the polydisperse Brownian rotation model and the equilibrium model is qualitatively similar to the relationship between their monodisperse versions.
The polydisperse N\'{e}el relaxation model changed its behavior and is now dominated by particles with smaller diameters resulting in slower changes in magnetic moment with a shorter delay.

\begin{table}

 \begin{tabular}{ll|l}
 \textbf{Parameter} & & \textbf{Value}\\ 

 \midrule
 Magnetic permeability & $\mu_0$ & $4\pi\times 10^{-7} \text{ H/m}$\\
 Boltzmann constant & $\kappa_\mathrm{B}$ & $1.38064852\times 10^{-23} \text{ J/K}$\\

\midrule 
\multicolumn{3}{l}{{\it Scanner}} \\
\midrule
 Excitation frequency & $f$& $25000 \text{ Hz}$\\
 Excitation amplitudes & $A_\mathrm{x}$& $0.012\text{ T}$\\
 Excitation repetition time & $T_\mathrm{R}$ & $0.04 \times 10^{-3} \text{ s}$\\
 Gradient strength & $G_{3,3}$ & $2\text{ T/m}$\\
 \midrule
\multicolumn{3}{l}{{\it Particle}} \\
 \midrule
  Temperature & $T_\textrm{B}$ & $300 \text{ K}$ \\
  Sat. magnetization & $M_\mathrm{C}$ & $474000 \text{ J/m$^3$/T}$\\
  Particle core diameter & $D$ & $20 \times 10^{-9} \text{ m}$ \\
  Particle core volume & $V_\mathrm{C}$ & $1/6 \pi D^3$ \\
  Particle hydrodynamic volume & $V_\mathrm{H}$ & $V_\mathrm{C}$ \\
  Dynamic viscosity (water) & $\eta$ & $1.0049 \times 10^{-3} \text{ Pa s} $\\
  Gyromagnetic ratio & $\gamma$ & $1.75 \times 10^{11} \text{ rad/s}$  \\
  Damping parameter & $\alpha$ & $0.1$ \\
  Anisotropy axis  & $n$ & $e_3$ \\
  Anisotropy constant & $K_\mathrm{anis}$ & $12000 \text{ J/m$^3$}$ \\
 
  \midrule
\multicolumn{3}{l}{{\it Tracer}} \\
 \midrule
   Mean lognormal distribution & $\mu$ & $\mathrm{ln}(13 \times 10^{-9})$ \\
  Standard deviation lognormal distribution & $\sigma$ & 0.37 \\
 \end{tabular}
 
 \caption{Physical parameters used for the simulations. The parameters can be found in: scanner setup \cite{Knopp2012r}, particle parameters \cite{Deissler2014}, and tracer distribution \cite{Eberbeck2011}.}
\label{tab:sim-parameters}
 \end{table}

\section{Discussion}
\label{sec:discussion}

We summarized several models relevant for magnetic particle imaging and showed simulations for one particular special case to illustrate the different behavior of the models.
These differences might be one of the reasons why model-based reconstructions with the equilibrium model are not of the same quality compared to reconstructions with a measured linear operator. 
The relaxation effects are considered independent of each other as it is commonly assumed that one cause for the relaxation dominates.
However, combined models considering Brownian rotation and N\'{e}el relaxation simultaneously are desirable and were investigated by using their Langevin equations \cite{Reeves2015combined}.

Imaging quality can suffer when neglecting particle relaxation as it modifies the measured time signal. 
A loss of quality is then due to the spatial encoding in the time-dependent signal. 
A certain delay like in Brownian rotation and N\'{e}el relaxation may cause shifted reconstructions in space when taking Cartesian trajectories into account. 
The damping and smoothing observed in the Brownian rotation model may cause an underestimated concentration and a spatially blurred reconstruction. 
The ill-posed nature of the problem allows for a certain degree of model errors only. 
In case of multidimensional trajectories, like Lissajous trajectories, the rotation of the applied magnetic field vector has to be taken into account.
Due to the loss of circular symmetry, the probability density function then needs to be approximated on the whole surface of the sphere.
The higher dimensionality also increases the computational costs such that more efficient numerical solutions are required to compute a solution of the probability density function. 
Furthermore, finding a direct and more efficient solution to compute the mean magnetic moment vector is highly desirable.  
Given the numerical solutions for the multidimensional case of the Brownian rotation and N\'{e}el relaxation, these models still need to be physically validated for applied magnetic fields in MPI.
For this purpose a recently proposed magnetic particle spectrometer (MPS) \cite{chen2017first} can provide the required measurements. 
The advantage of the proposed MPS is that it allows applying a drive field with a three-dimensional excitation pattern and a constant offset field simulating the selection field at a fixed position.

This overview about mathematical models for magnetic particle imaging builds the basis for several directions of future research. 
The numerical treatment of all models in the multidimensional case requires further analysis and the development of efficient algorithms. 
A first step into the direction of theoretical investigations was made by formulating a different problem setting motivated by the equilibrium model \cite{maerz2015modelbased}. 
But by neglecting the temporal dependencies in the methodology, the equilibrium model as defined in this work is not directly covered.
The mathematical models for MPI summarized in this work have not been investigated analytically including the related inverse problems.
Besides the linear inverse problem of reconstructing the concentration, several nonlinear inverse problems are motivated by applications and the particle behavior itself.
For example, joint concentration reconstructions combined with a spatial viscosity or temperature distribution were already motivated in the context of multi-color MPI \cite{rahmer2015first,stehning2016simultaneous}.
Interactions between particles cause a nonlinear concentration dependence which becomes of interest \cite{loewa2016}.
Prior the consideration of combined problems, a series of analytical works regarding the presented models is required.

\section*{Acknowledgements}
T. Kluth is supported by the Deutsche Forschungsgemeinschaft (DFG) within the framework of GRK 2224/1 ``Pi$^3$ : Parameter Identification - Analysis, Algorithms, Applications''. 
The author also thanks Bangti Jin, University College London, for his helpful comments on this manuscript and fruitful discussions on MPI. 
\vspace{5mm}

\bibliographystyle{abbrv}
\bibliography{literature,ref,ref2}

\appendix

\section{Fokker-Planck equation for Brownian rotation}
\label{app:FP-Brown}
We derive \myEqref{eq:problem-mono-brown} from the Langevin equation of the particle dynamics.
As the spatial dependence in the probability density function $f$ solely results implicitly from the spatially dependent applied magnetic field $B_\mathrm{app}$, we consider the problem of determining $f$ for one fixed $x \in  \Omega$, and further omit the spatial variable $x$.
The Langevin equation for Brownian rotation of a single particle with magnetic moment vector $\tilde{m}$ reads
\begin{equation}
 \frac{\partial}{\partial t} \tilde{m} = \frac{1}{6 V_\mathrm{H} \eta} \left( \tilde{m} \times B_\mathrm{eff} + \tilde{D} \Gamma \right) \times \tilde{m},
\end{equation}
with $\tilde{D}>0$ determined below and where $\Gamma$ is a white noise component with $\langle \Gamma_i(t)\rangle =0$, $\langle \Gamma_i(t_1) \Gamma_j(t_2) \rangle = \delta_{ij} \delta(t_1-t_2)$, for all $t,t_1,t_2>0$ and $i,j=1,2,3$.
$\delta_{i,j}$ is the Kronecker delta and $\delta$ is the Dirac delta distribution.
Substituting $m=\tilde{m}/m_0$, $m_0=|\tilde{m}|$ yields 
\begin{align}
 \frac{\partial}{\partial t} m &= \frac{1}{6 V_\mathrm{H} \eta} \left( m \times m_0 B_\mathrm{eff} + \tilde{D} \Gamma \right) \times m \notag \\
 &= \frac{1}{6 V_\mathrm{H} \eta} \left( m_0( m \times  B_\mathrm{eff})\times m + \tilde{D} \Lambda(m)\Gamma \right) 
\end{align}
where 
\begin{equation}
 \Lambda(m)= \begin{pmatrix} 0 & -m_3 & m_2 \\ m_3 & 0 & -m_1\\ -m_2 & m_1 & 0 \end{pmatrix}.
\end{equation}
The Fokker Planck equation for the probability density function $f:S\times (0,T) \rightarrow \R_+$ thus becomes \cite{risken1996fokker} 
\begin{equation}
 \frac{\partial}{\partial t} f (m,t) = - \mathrm{div}_m \left( a(m,t) f(m,t) + \frac{1}{2} B(m,t) \left(\mathrm{div}_m(B^{(j)}(m,t) f(m,t))\right)_{j=1,2,3}   \right)
\end{equation}
where $a: S \times (0,T) \rightarrow \R^3$ and $B: S \times (0,T) \rightarrow \R^{3\times 3}$ with $B^{(j)}$ being the $j$-th column of $B$. 
From the Langevin equation it follows
\begin{equation}
 a(m,t) = \frac{m_0}{6 V_\mathrm{H} \eta} \left( B_\mathrm{eff} - (m\cdot B_\mathrm{eff}) m \right)
\end{equation}
and 
\begin{equation} 
 B(m,t)=\frac{\tilde{D}}{6 V_\mathrm{H} \eta} \Lambda(m)^T.
\end{equation}
By using the fact that $m\cdot \nabla_m f=0$ (as the gradient is tangent to the unit sphere) we obtain by using $\Lambda(m)^T \Lambda(m) \nabla_m f = ( m \times \nabla_m f ) \times m=\nabla_m f$ the desired Focker Planck equation
\begin{equation}
\frac{\partial } {\partial t} f(m,t) = - \frac{1}{6V_\mathrm{H} \eta} \mathrm{div}_m \left( m_0 ( B_\mathrm{eff} - (m\cdot B_\mathrm{eff}) m) f(m,t) - \frac{\tilde{D}^2}{12 V_\mathrm{H} \eta} \nabla_m f(m,t) \right)
\end{equation}
The diffusion coefficient $\tilde{D}$ is determined by considering the equilibrium case \cite{rogge2013simulation}, i.e. let $t_0 \in I$ be the point in time such that $\frac{\partial}{\partial t} f(m,t_0) =0$.
We further assume that  $f_0: S \rightarrow \R^+ \cup \{0\}$ with $f_0(\cdot) = f(\cdot,t_0)$ corresponds to the Boltzmann distribution for Brownian rotation in the equilibrium, i.e., $f_0(m) = k e^{-\beta \mathcal{H}(m,t_0)}$, where $\mathcal{H}$ is the Hamiltonian with $\nabla_m \mathcal{H}=-B_\mathrm{eff}$ and $\beta = \frac{m_0}{\kappa_\mathrm{B} T_\mathrm{B}}$.
Assume $B_\mathrm{eff}$ does not depend on $m$, which holds if $B_\mathrm{eff}=B_\mathrm{app}$. 
We use $\mathcal{H}(m,t) = - B_\mathrm{app} \cdot m $ and $\nabla_m f_0 \cdot m =0$ to obtain 
\begin{align}
0 &=  \mathrm{div}_m \left( m_0 ( B_\mathrm{app} - (m\cdot B_\mathrm{app}) m) e^{ \frac{m_0}{\kappa_\mathrm{B} T_\mathrm{B}} B_\mathrm{app} \cdot m}- \frac{\tilde{D}^2}{12 V_\mathrm{H} \eta} \frac{m_0}{\kappa_\mathrm{B} T_\mathrm{B}} B_\mathrm{eff} e^{ \frac{m_0}{\kappa_\mathrm{B} T_\mathrm{B}} B_\mathrm{app} \cdot m} \right)\notag \\
&= \frac{m_0^2}{\kappa_\mathrm{B} T_\mathrm{B}} |B_\mathrm{app}|^2 - \frac{\tilde{D}^2}{12 V_\mathrm{H} \eta} \left(\frac{m_0}{\kappa_\mathrm{B} T_\mathrm{B}}\right)^2 |B_\mathrm{app}|^2.
\end{align}
From this it follows $\tilde{D}= \sqrt{12 V_\mathrm{H} \eta \kappa_\mathrm{B} T_\mathrm{B}}$. 
Defining $\tau= \frac{3 V_\mathrm{H} \eta }{\kappa_\mathrm{B} T_\mathrm{B}}$ yields
\begin{equation}
\frac{\partial } {\partial t} f(m,t) = - \frac{1}{2 \tau} \mathrm{div}_m \left( \frac{m_0}{\kappa_\mathrm{B} T_\mathrm{B}} ( B_\mathrm{app} - (m\cdot B_\mathrm{app}) m) f(m,t) -  \nabla_m f(m,t) \right).
\end{equation}

\section{Derivation of Langevin function} 
\label{app:FP-Equilibrium}
In the following we give an example how the equilibrium model and particularly using the Langevin function is motivated and related to Brownian rotation.
As can be seen in \ref{app:FP-Brown}, the Fokker-Planck equation is parametrized such that the probability density function of the magnetic moment vector in equilibrium at time $t_0\in I$ is of the form
\begin{equation}
 f_0(m)= k e^{-\beta \mathcal{H}(m,t_0)}.
\end{equation}
Considering the mean magnetic moment vector yields
\begin{align}
 \bar{m}(t_0) & = m_0 \int_S m f_0(m) \ dm 
=m_0 k \int_S m e^{\beta (R^Tm)\cdot (R^T  B_\mathrm{app}(t_0))} \ dm \notag  \\ 
& = m_0 k R \int_S y e^{y_3 \beta b_0} \ dy \notag \\
& = m_0 k R  \int_0^\pi  \int_{0}^{2\pi} \begin{pmatrix} \sin(\theta) \cos(\phi) \\ \sin(\theta) \sin(\phi) \\ \cos(\theta) \end{pmatrix} |\sin(\theta)|  e^{\cos(\theta) \beta b_0} \ d \phi \ d\theta \notag  \\
&= 2 \pi m_0 k R e_3 \int_0^\pi \cos(\theta) \sin(\theta)  e^{\cos(\theta) \beta b_0} \ d\theta \notag \\
&= -2 \pi m_0 k R e_3 \int_{-1}^{1} x e^{x \beta b_0} \ dx \notag \\
& = -2 \pi m_0 k  R e_3 \left( \left[\frac{1}{\beta b_0} x e^{x \beta b_0}\right]_{-1}^1 - \int_{-1}^1 \frac{1}{\beta b_0} e^{x \beta b_0} \ dx   \right) \notag \\
&=- R e_3 m_0 \left( \frac{e^{\beta b_0} + e^{-\beta b_0}}{e^{\beta b_0} - e^{-\beta b_0}} - \frac{1}{\beta b_0} \right)  
\end{align}
where $R\in\R^{3\times 3}$ is a rotation matrix sucht that $R^T B_\mathrm{app}(t_0)=e_3 b_0$, where $b_0=|B_\mathrm{app}(t_0)|$, and where $k$ is obtained by similar transformations such that it is given by
\begin{equation}
\frac{1}{k}= \int_S e^{-\beta \mathcal{H}(m,t)} \ d m = \int_S e^{y_3 \beta  b_0} \ d y = 2 \pi  \int_{-1}^1 e^{x \beta b_0} \ dx =  \frac{2 \pi}{\beta b_0} ( e^{\beta b_0} - e^{- \beta b_0}).
\end{equation}
The length of the mean magnetic moment is thus given by $|\bar{m}(t_0)|= \mathcal{L}_\beta( |B_\mathrm{app}(t_0)|)$ with $\mathcal{L}_\beta$ given by \myEqref{eq:Langevin}.

\section{Fokker-Planck equation for N\'{e}el relaxation}
\label{app:FP-Neel}
We derive \myEqref{eq:problem-mono-neel} from the Langevin equation based on the Landau-Lifshitz-Gilbert equation for the particle dynamics.
As the spatial dependence in the probability density function $f$ solely results implicitly from the spatial dependent applied magnetic field $B_\mathrm{app}$ we consider the problem of determining $f$ for one fixed $x \in  \Omega$, 
and further omit the spatial variable $x$.
The Langevin equation for N\'{e}el relaxation of a single particle with magnetic moment vector $\tilde{m}$ is obtained from the Landau-Lifshitz-Gilbert equation and reads
\begin{equation}
 \frac{\partial}{\partial t} \tilde{m} = \tilde{\gamma}  \left( (B_\mathrm{eff} + \tilde{D} \Gamma ) \times \tilde{m} + \frac{\alpha}{m_0} (\tilde{m} \times (B_\mathrm{eff} + \tilde{D} \Gamma) ) \times \tilde{m}\right),
\end{equation}
with $\tilde{D}>0$ determined below and where $\tilde{\gamma}=\gamma/(1+\alpha^2)$ and where $\Gamma$ is a white noise component with $\langle \Gamma_i(t)\rangle =0$, $\langle \Gamma_i(t_1) \Gamma_j(t_2) \rangle = \delta_{ij} \delta(t_1-t_2)$, for all $t,t_1,t_2>0$ and $i,j=1,2,3$.
$\delta_{i,j}$ is the Kronecker delta and $\delta$ is the Dirac delta distribution.
Substituting $m=\tilde{m}/m_0$, $m_0=|\tilde{m}|$ yields 
\begin{align}
 \frac{\partial}{\partial t} m &= \tilde{\gamma}  \left( (B_\mathrm{eff} + \tilde{D} \Gamma ) \times m + \alpha (m \times (B_\mathrm{eff} + \tilde{D} \Gamma )) \times m\right)\notag \\
 &= \tilde{\gamma}  \left( B_\mathrm{eff}  \times m + \alpha  (m \times  B_\mathrm{eff} ) \times m + \tilde{D} (\Lambda(m)^T+\alpha  \Lambda(m)^T \Lambda(m) ) \Gamma \right) 
\end{align}
where 
\begin{align}
 \Lambda(m)&= \begin{pmatrix} 0 & -m_3 & m_2 \\ m_3 & 0 & -m_1\\ -m_2 & m_1 & 0 \end{pmatrix} \\ \land \  \Lambda(m)^T \Lambda(m) &= -\begin{pmatrix} m_2^2+m_3^2 & -m_1 m_2 & -m_1 m_3 \\ -m_2 m_1 & m_1^2+m_3^2 & -m_2 m_3\\ -m_3 m_1 & -m_3 m_2 & m_1^2 + m_2^2 \end{pmatrix}.
\end{align}
The Fokker-Planck equation for the probability density function $f:S\times (0,T) \rightarrow \R^+ \cup \{0\}$ thus becomes \cite{risken1996fokker} 
\begin{equation}
 \frac{\partial}{\partial t} f (m,t) = - \mathrm{div}_m \left( a(m,t) f(m,t) + \frac{1}{2}  B(m,t) \left(\mathrm{div}_m(B^{(j)}(m,t) f(m,t))\right)_{j=1,2,3}   \right)
\end{equation}
where $a: S \times (0,T) \rightarrow \R^3$ and $B: S \times (0,T) \rightarrow \R^{3\times 3}$ with $B^{(j)}$ being the $j$-th column of $B$. 
From the Langevin equation it follows
\begin{equation}
 a(m,t) = \tilde{\gamma}  \left( B_\mathrm{eff}  \times m + \alpha  (m \times  B_\mathrm{eff} ) \times m  \right)
\end{equation}
and 
\begin{equation} 
 B(m,t)=\tilde{D} \tilde{\gamma}\left(\Lambda(m)^T+\alpha \Lambda(m)^T \Lambda(m) \right).
\end{equation}
By using the fact that $m\cdot \nabla_m f=0$ (as the gradient is tangent to the unit sphere) we obtain by using $\Lambda(m)^T \Lambda(m) \nabla_m f = ( m \times \nabla_m f ) \times m =\nabla_m f$ the desired Focker Planck equation
\begin{align}
&\frac{\partial } {\partial t} f(m,t) \notag \\ =& - \tilde{\gamma} \mathrm{div}_m \left(\left(B_\mathrm{eff} \times m +  \alpha  ( B_\mathrm{eff} - (m\cdot B_\mathrm{eff}) m)\right) f(m,t) - \frac{\tilde{D}^2 \tilde{\gamma} (1+ \alpha^2 )}{2} \nabla_m f(m,t) \right)
\end{align}
The diffusion coefficient $\tilde{D}$ is determined by considering the equilibrium case \cite{rogge2013simulation}, i.e. let $t_0 \in I$ be the point in time such that $\frac{\partial}{\partial t} f(m,t_0) =0$.
We further assume that  $f_0: S \rightarrow \R^+ \cup \{0\}$ with $f_0(\cdot) = f(\cdot,t_0)$ corresponds to the Boltzmann distribution for Brownian rotation in the equilibrium, i.e., $f_0 = k e^{-\beta \mathcal{H}(m,t_0)}$, where $\mathcal{H}$ is the Hamiltonian with $\nabla_m \mathcal{H}=-B_\mathrm{eff}$ and $\beta = \frac{m_0}{\kappa_\mathrm{B} T_\mathrm{B}}$.
We use $\nabla_m f_0 =  \beta B_\mathrm{eff} f_0 $ and $\nabla_m f_0 \cdot m =0$ to obtain 
\begin{align}
0 &=  \mathrm{div}_m \left(\left(B_\mathrm{eff} \times m +  \alpha  ( B_\mathrm{eff} - (m\cdot B_\mathrm{eff}) m)\right) f_0 - \frac{\tilde{D}^2 \tilde{\gamma} (1+ \alpha^2 )}{2} \nabla_m f_0 \right)\notag \\
&=  \mathrm{div}_m \left((B_\mathrm{eff} \times m) f_0\right) + \alpha  \mathrm{div}_m \left(  B_\mathrm{eff} f_0 \right)  - \frac{\tilde{D}^2 \tilde{\gamma} (1+ \alpha^2 )}{2} \Delta_m f_0 \notag \\
&=  \mathrm{div}_m \left(B_\mathrm{eff} \times m\right) f_0 + \left( \frac{\alpha}{\beta}  - \frac{\tilde{D}^2 \tilde{\gamma} (1+ \alpha^2 )}{2} \right) \Delta_m f_0 \notag \\
&=  \left( \frac{\alpha }{\beta}  - \frac{\tilde{D}^2 \tilde{\gamma} (1+ \alpha^2 )}{2} \right) \Delta_m f_0
\end{align}
where the last inequality holds as $\mathrm{div}_m \left(B_\mathrm{eff} \times m\right)=0$ for $B_\mathrm{eff}= B_\mathrm{app} + B_\mathrm{anis}$.
It follows $\tilde{D}= \sqrt{2 \frac{\alpha\kappa_\mathrm{B} T_\mathrm{B}}{\tilde{\gamma} (1+\alpha^2) m_0}} $. 
Defining $\tau= \frac{m_0}{2 \alpha \tilde{\gamma} \kappa_\mathrm{B} T_\mathrm{B}}$ yields
\begin{equation}
\frac{\partial } {\partial t} f(m,t) = - \mathrm{div}_m \left( \tilde{\gamma} \left( B_\mathrm{eff} \times m + \alpha ( B_\mathrm{eff} - (m\cdot B_\mathrm{eff}) m) \right) f(m,t) - \frac{1}{2 \tau}  \nabla_m f(m,t) \right).
\end{equation}

\section{Numerical solution for 1D excitation}
\label{app:FP-numerics1D}

The following numerical solution follows \cite{Deissler2014}.
We compute the mean magnetic moment vector $\bar{m}$ for an applied magnetic field of the form $B_\mathrm{app}(x,t) = b(x,t)e_3$
with a given field $b: \Omega \times I \rightarrow \R$. 
We additionally assume that $\Omega \subset \{t e_3 | t\in \R\}$.

Under these assumptions the Fokker-Planck equations for Brownian rotation in \myEqref{eq:problem-mono-brown} and N\'{e}el relaxation in \myEqref{eq:problem-mono-neel} can be reformulated in terms of the angle $\theta$ between $m$ and $e_3$. 
Let $x\in \Omega$ be fixed.
By assuming $z=\cos(\theta)$ the resulting ordinary differential equations for $\tilde{f}_x: [-1,1] \times I  \rightarrow \R^+ \cup \{ 0 \}$ then become for Brownian rotation
\begin{equation}
 2 \tau_\mathrm{B} \frac{\partial}{\partial t} \tilde{f}_x(z,t) = \frac{\partial}{\partial z} \left( (1-z^2) \left( \frac{\partial}{\partial z} \tilde{f}_x(z,t) - \beta b(x,t) \tilde{f}_x(z,t) \right) \right) 
\end{equation}
with physical parameters $\tau_\mathrm{B}=3 V_\mathrm{H} \eta / (\kappa_\mathrm{B} T_\mathrm{B})$ and $\beta=\frac{m_0}{\kappa_\mathrm{B} T_\mathrm{B}}$. For N\'{e}el relaxation we obtain
\begin{equation}
 2 \tau_\mathrm{N} \frac{\partial}{\partial t} \tilde{f}_x(z,t) = \frac{\partial}{\partial z} \left( (1-z^2) \left( \frac{\partial}{\partial z} \tilde{f}_x(z,t) - \beta b(x,t) \tilde{f}_x(z,t) - \tilde{k} z \tilde{f}_x(z,t) \right) \right) 
\end{equation}
with physical parameters $\tau_\mathrm{N}=\frac{m_0}{2 \alpha \tilde{\gamma} \kappa_\mathrm{B} T_\mathrm{B}}$, $\beta=\frac{m_0}{\kappa_\mathrm{B} T_\mathrm{B}}$, and $\tilde{k}=\frac{2 K_\mathrm{anis} m_0}{M_\mathrm{C} \kappa_\mathrm{B} T_\mathrm{B}} $.
The mean magnetic moment vector $\bar{m}$ is then obtained by 
\begin{equation}
 \bar{m}(x,t) = m_0 \int_{-1}^1 z \tilde{f}_x(z,t) \ dz  e_3.
\end{equation}
We expand the probability density function $\tilde{f}$ in Legendre polynomials $\{P_n\}_{n\in \N}$ and $a_n: I \rightarrow \R$, $n \in \N$, i.e.,
\begin{equation}
 \tilde{f}(z,t) = \sum_{n=0}^\infty a_n(t) P_n(z).
\end{equation}
Using this relation and the constraint $\int_{-1}^1 \tilde{f}(z,t) \ dz =1$, $t \in I$, we obtain the system of differential equations for Brownian rotation
\begin{equation}
\left\{
\begin{split}
 a_0(t) &= \frac{1}{2} & & \\
 \frac{2 \tau_\mathrm{B} }{n(n+1)} \frac{d}{dt} a_n(t) &= - a_n(t) + \beta b(x,t) \left(\frac{a_{n-1}(t)}{2n-1} - \frac{a_{n+1}}{2n+3} \right) & & n\geq 1   
\end{split}
\right.
\end{equation}
and for N\'{e}el relaxation
\begin{equation}
\left\{
\begin{split}
 a_0(t) &= \frac{1}{2} & & \\
 \frac{2 \tau_\mathrm{N} }{n(n+1)} \frac{d}{dt} a_n(t) &= - a_n(t) + \beta b(x,t) \left(\frac{a_{n-1}(t)}{2n-1} - \frac{a_{n+1}}{2n+3} \right) & & \\ 
&+ \tilde{K} \left( \frac{(n-1) a_{n-2}(t)}{(2n-3)(2n-1)} + \frac{n a_{n}(t)}{(2n-1)(2n+1)} \right. & & \\
& \left. - \frac{(n+1) a_{n}(t)}{(2n+1)(2n+3)}- \frac{(n+2) a_{n+2}(t)}{(2n+3)(2n+5)}\right)& & n\geq 1   
\end{split}
\right.
\end{equation}
 by assuming an $a_{-1}(t) = 0$. 
Assuming a uniform distribution as initial condition results in the initial values $a_0(0)=1/2$ and $a_n(0)=0$, $n\geq 1$, for both systems. 
Using the relation $\int_{-1}^1 z \tilde{f}_x(z,t) \ dz = 2/3 a_1(t)$ determines the mean magnetic moment vector. 
The time derivative of the mean magnetic moment vector is thus given by the respective differential equation for $n=1$. 

For the simulation we use an approximated probability density function for $N\in\N$ given by 
\begin{equation}
 \tilde{f}(z,t) = \sum_{n=0}^N a_n(t) P_n(z).
\end{equation}
The applied magnetic field is given by 
\begin{equation}
 b(x,t)= A \cos(2 \pi f t) + G_{3,3} x_3.
\end{equation}
where $A,f >0$ and $G_{3,3}$. 
The physical parameters used for the simulation can be found in Table \ref{tab:sim-parameters}.
\end{document}